\documentclass[12pt,preprint]{revtex4}
\usepackage{graphicx}
\usepackage{amsmath}
\usepackage{hyperref}
\usepackage{bm}
\usepackage[dvips]{color}
\usepackage{sidecap}

\graphicspath{{figs/},
      }

\newcommand{\degr}{$^{\circ}$}

\newcommand{\fzd} {Institute of Ion Beam Physics and Materials Research, Forschungszentrum Dresden-Rossendorf, P.O. Box 510119, 01314 Dresden, Germany}

\begin{document}
\title{Anomalous Hall resistance in Ge:Mn systems with low Mn concentrations}
\date{\today}

\author{Shengqiang~Zhou}
\email[Electronic address: ]{S.Zhou@fzd.de} \affiliation{\fzd}
\author{Danilo~B\"{u}rger}
\author{Manfred~Helm}
\affiliation{\fzd}
\author{Heidemarie~Schmidt}
\affiliation{\fzd}

\begin{abstract}
Taking Mn doped Germanium as an example, we evoke the consideration of a two-band-like conduction in diluted ferromagnetic semiconductor (FMS).
The main argument for claiming Ge:Mn as a FMS is the occurrence of the anomalous Hall effect (AHE). Usually, the reported AHE (1) is observable
at temperatures above 10 K, (2) exhibits no hysteresis, and (3) changes the sign of slope. We observed a similar Hall resistance in Mn implanted
Ge with the Mn concentration as low as 0.004\%. We show that the puzzling AHE features can be explained by considering a two-band-like
conduction in Ge:Mn.

\end{abstract}
\maketitle

Diluted ferromagnetic semiconductors (FMS) exhibit strong magneto-transport effects, namely negative magnetoresistance (MR) and anomalous Hall
effect (AHE) \cite{hayashi:4865,PhysRevLett.68.2664,PhysRevB.63.085201}, and provide the possibility to control the spin by an external electric
field. Ferromagnetic GaMnAs reveals hysteretic AHE, which mimics its magnetization, and allows the determination of its magnetic parameters by
measuring the Hall resistance. The observation of AHE is considered as one of the important criteria for FMS to be intrinsic \cite{ohnoGaMnAs}.
Mn doped Germanium provides an alternative of FMS, as predicted by Dietl \emph{et al.} \cite{dietl00}. Its compatibility with conventional
microelectronics makes it more promising for industry application. We notice that pronounced MR and AHE have been reported in the Ge:Mn system
\cite{Park01252002,PhysRevLett.91.177203,jamet06,PhysRevB.72.195205,riss:241202,zeng:066101,gareev:222508,deng:062513} independent of the
formation of MnGe precipitates or not, as well as in Cr doped Ge \cite{1742-6596-51-1-025}. By scrutinizing the published data on Ge:Mn, one can
observe three features in the reported AHE. First, most of the AHE curves shown were recorded at temperatures above 10 K. Indeed, Riss \emph{et
al.} \cite{riss:241202} reveal only ordinary Hall effect below 10 K. Second, no hysteresis in AHE curves has been observed, despite the
observation of a clear hysteresis in magnetization, which is much different from the case of III-Mn-V and ZnMnTe
\cite{hayashi:4865,PhysRevLett.68.2664,PhysRevB.63.085201}. Third, the Hall curve changes the sign of slope at lower temperatures, usually
between 10 K and 50 K. Obviously, the correlation between magnetization, MR, and AHE, which is a hallmark of III-Mn-V and ZnMnTe FMS
\cite{hayashi:4865,PhysRevLett.68.2664,PhysRevB.63.085201}, has not been proven for Ge:Mn. In this Letter, we report the observation of an
anomalous Hall resistance in Mn implanted Ge with the Mn concentration as low as 0.004\%, in which neither ferromagnetism nor paramagnetism has
been measured. By considering two types of carriers, all the puzzling Hall-resistance features can be explained. Moreover, the two-band-like
conduction probably also explains the Hall resistance at large fields for InMnAs \cite{PhysRevB.59.5826}  and InMnSb \cite{mihly:107201}.

Intrinsic Ge(001) wafers were implanted with Mn ions at 300 {\degr}C to avoid amorphization. We varied the ion fluence to get samples with a
large range of Mn concentrations and correspondingly different structural and magnetic properties (see Tab. \ref{tab:sample}). Magnetotransport
properties were measured using Van der Pauw geometry with a magnetic field applied perpendicular to the film plane.

As shown in Tab. \ref{tab:sample}, we examined samples with a wide range of Mn concentration. Mn$_5$Ge$_3$ precipitates have been observed in
sample Ge02 and Ge03 by synchrotron radiation X-ray diffraction (SR-XRD) \cite{zhou_Mn5Ge3} at the Rossendorf beamline (BM20) at the ESRF, but
not in sample Ge01, and are not expected in sample Ge19. SR-XRD has been proved to be sensitive for detecting nanocrystalline precipitates
\cite{pot06fe,zhou07si}. Correspondingly, ferromagnetism was observed only in sample Ge02 and Ge03. Down to 5 K, only diamagnetism was probed
for sample Ge01, identical to a virgin Ge sample. Note that independent of the formation of precipitates, a fraction of Mn ions has been
confirmed by spectroscopic methods to be diluted inside the Ge matrix, resulting in p-type doping
\cite{picozzi:062501,gambardella:125211,biegger:103912}, as well as by electrical transport measurements as shown in Figure 1(a). In sharp
contrast to the structural and magnetic properties, similar Hall effects are probed for all samples as shown in Figure 1. Indeed, the shape and
temperature dependence of the Hall-resistance curves are similar to those reported in literature
\cite{PhysRevLett.91.177203,jamet06,PhysRevB.72.195205,riss:241202,zeng:066101,gareev:222508,deng:062513}. Deng \emph{et al.} attribute the sign
change in Hall curves to the hopping conductivity \cite{deng:062513}. However, this cannot explain the vanishing of AHE below 10 K.

The non-ferromagnetic nature of sample Ge19 and Ge01 indicates that the observation of anomalous Hall resistance is not necessarily related to
ferromagnetism. Actually, similar Hall curves have been observed in materials with a two-band-like conduction
\cite{PhysRevB.42.3578,PhysRevB.61.9621,jung:043703}. Figure 2 shows the resistance vs. temperature for all samples. Basically, below 10 K the
resistance is weakly dependent on temperature and a transition occurs above 10 K. The activation energy is only several meV for all samples. It
can not be the thermal activation of holes from the Mn acceptor levels in Ge since the Mn single and double acceptor level is as deep as 160 meV
from the valence band and 370 meV from the conduction band \cite{PhysRev.100.659}, respectively. The small activation energy also cannot be a
shallow acceptor level due to the large difference in the magnitude and temperature-dependence of resistance compared to Ga or In-doped Ge
\cite{PhysRev.99.406}. Moreover, between 10 K and 50 K the temperature dependence of the resistance is rather a universal feature in Mn doped Ge
\cite{Park01252002,park:2739,PhysRevLett.91.177203,PhysRevB.72.165203,riss:241202}. An interpretation of the transition at around 10 K is
plausible by considering the ground and the first excited states of Mn$^{2+} ($$d^5+2\mathrm{h}$) \cite{ZhouMnGe}. Within such a scenario, a
two-band-like conduction (or two types of carriers) is expected at above 10 K or even lower if the Mn concentration is very small as shown in
Figure 2. In the next step we fit the Hall curves using a two-band model described in Refs. \onlinecite{PhysRevB.61.9621} and
\onlinecite{jung:043703}.

The dependence of the ratio between Hall and sheet resistance (R$_{xy}$/R$_{xx}$) on the magnetic field B is given by
\begin{equation}\label{Rhall}
    R_{xy}/R_{xx}=\frac{x_1\mu_1(1+\mu_1^2B^2)+x_2\mu_2[1+\mu_2^2B^2]}{1+(x_1\mu_2+x_2\mu_1)^2B^2}B,
\end{equation}
where \begin{equation}\label{xx}
    x_1=\frac{G_1}{G_1+G_2},~x_2=\frac{G_2}{G_1+G_2},
\end{equation} $G$ and $\mu$ are conductance and mobility, respectively. The subscript numbers denote the two types of carriers. Numbers '1' and '2' label the carriers in
the ground and the first excited state, respectively. In this formula, we assume spherical Fermi surfaces and use the zero-field sheet
resistance ($R_{xx}$) to relate the two-carrier conductance \cite{PhysRevB.61.9621},
\begin{equation}\label{}
    1/R_{xx}=G=G_1+G_2.
\end{equation}
$x_1$, $\mu_1$ and $\mu_2$ in Eq. (\ref{Rhall}) are fitting parameters. At 5 K, for sample Ge01, Ge02, and Ge03 we only have carriers of type-1.
With increasing temperature carriers of type-2 are activated and $x_1$ decreases. Taking this as a restriction, we can fit all measured results.
Figure 1(d) shows the comparison between experiments and fitting. All the features are reproduced. In Figure 3 we list the fitting results,
\emph{i.e.}, temperature dependent mobility and sheet carrier concentration for two types of carriers. Despite the number of fitting parameters,
it is plausible to draw conclusions according to the overall trend by examining all samples. The sheet carrier concentration (\emph{N}$_{H1}$)
of type-1 carriers is weakly dependent on temperature [Figure 3(c)], while \emph{N}$_{H2}$ is increased above 10 K and becomes saturated at high
temperatures [Figure 3(d)]. The dependence of $\mu_1$ on temperature is nonmonotonic. This is expected given the fact of a large hole
concentration in those samples. The ion implantation results in an effective Ge:Mn layer thickness of around 100 nm. Then the concentration of
type-1 carriers is in the range between 10$^{17}$ and 10$^{18}$ cm$^{-3}$. In this range a peak of hole mobility in Ge depending on temperature
was observed \cite{Golikova}. The concentration of type-2 carriers is mostly below 10$^{17}$ cm$^{-3}$ and $\mu_2$ increases monotonically with
decreasing temperature.

Eq. (\ref{Rhall}) neglects the influence of a possible paramagnetism inside the samples. The paramagnetism can be included as an anomalous Hall
term using a Brillouin function \cite{jung:043703}. However, we have not observed any paramagnetic component using SQUID magnetometry down to 5
K with fields up to 7 T for sample Ge19 and Ge01. The quench of the magnetic moments from Mn ions is somehow surprising, and has also been
observed in Mn doped Si \cite{zengli}. Nevertheless, we also refined Eq. (\ref{Rhall}) by including a Brillouin function in order to account for
possible magnetic field dependent mobilities. However, using the Brillouin function the fitting of the Hall curve at large fields is not
possible because the fitted Hall curve changes the sign of slope with increasing field. This is in contrast to the Hall data probed on Mn
implanted Ge where the sign of slope remains positive at large fields \cite{riss:241202}. The simple two-band-like picture described by Eq.
(\ref{Rhall}) also qualitatively explains the saturation of the positive MR at large fields. Namely, above 10 K a large positive MR is usually
observed in Ge:Mn (also in our samples) and saturates at large fields \cite{PhysRevLett.91.177203,park:2739,PhysRevB.72.195205,riss:241202},
which is a typical feature of the two-band MR \cite{PhysRevB.61.9621}.

In summary, we observed an anomalous Hall resistance in Mn implanted Ge with the Mn concentration as low as 0.004\%. By considering two types of
carriers participating in the conduction, we can explain the puzzling Hall-resistance curves reported in the literature and in this letter. A
multiple-path-like conduction can mislead the interpretation of materials as FMS. Possibly, it also can make an add-on effect to the AHE
observed in InMnAs \cite{PhysRevB.59.5826}  and InMnSb \cite{mihly:107201}, as well as in Zn$_{0.9}$Mn$_{0.075}$Cu$_{0.025}$O \cite{Xu20081160}.

The authors (S.Z. and D.B.) thank financial support from the Bundesministerium f\"{u}r Bildung und Forschung (FKZ13N10144). We thank Carsten
Timm from TU Dresden for fruitful discussion.


\clearpage

\begin{table}
\caption{\label{tab:sample} Sample identification (ID), Mn concentration (Mn conc.), sheet hole concentration (Hole conc.) and magnetic
properties.}
\begin{ruledtabular}
\begin{tabular}{ccccccc}
  ID  & Mn conc. &  Hole conc. & \multicolumn{2}{c} {Properties} \\
         &  \%  & cm$^{-2}$& Precipitates? & Ferromagnetic? \\
  \hline
Ge19 & 0.004 & - & No & No \\
Ge01 & 0.2 & 6.5$\times10^{12}$& No & No \\
Ge02 & 2 & 1.1$\times10^{13}$ & Mn$_{5}$Ge$_3$ & Yes \\
Ge03 & 10 & 2.0$\times10^{13}$ & Mn$_{5}$Ge$_3$ & Yes \\
\end{tabular}
\end{ruledtabular}
\end{table}

\clearpage Fig captions

Fig. 1 (a) Hall resistance (\emph{R}$_{xy}$) at 5 K: only ordinary Hall effect has been observed. (b) and (c): The ratio between \emph{R}$_{xy}$
and sheet resistance at zero field (\emph{R}$_{xx}$) at 20 K and 50 K, respectively. An anomalous Hall resistance appears and the sign of slope
is changed at lower fields (20 K) or at larger fields (50 K). (d) \emph{R}$_{xy}$/\emph{R}$_{xx}$ at different temperatures for sample Ge03: the
symbols are experimental data, while the solid lines are fits using Eq. (\ref{Rhall}). A Ga-doped Ge wafer with a hole concentration of
1.5$\times10^{16}$ cm$^{-3}$ was measured for comparison and only ordinary Hall effect is observed as shown in (c). For better visibility some
curves are multiplied by the factors indicated.

Fig. 2 Temperature dependent sheet resistance at zero field (\emph{R}$_{xx}$). For sample Ge19 the resistance is too large and can only be
measured down to 10 K. Basically, below 10 K the resistance of sample Ge02 and Ge03 is weakly dependent on temperature.

Fig. 3 Fitting results of Hall curves at different temperatures. (a) and (b): Temperature dependent mobilities, data extracted from Ref.
\onlinecite{Golikova}~is shown for comparison. (c) and (d): Temperature dependent sheet carrier concentrations. Lines are guides for eyes.

\clearpage

\begin{figure*} \center
\includegraphics[scale=0.8]{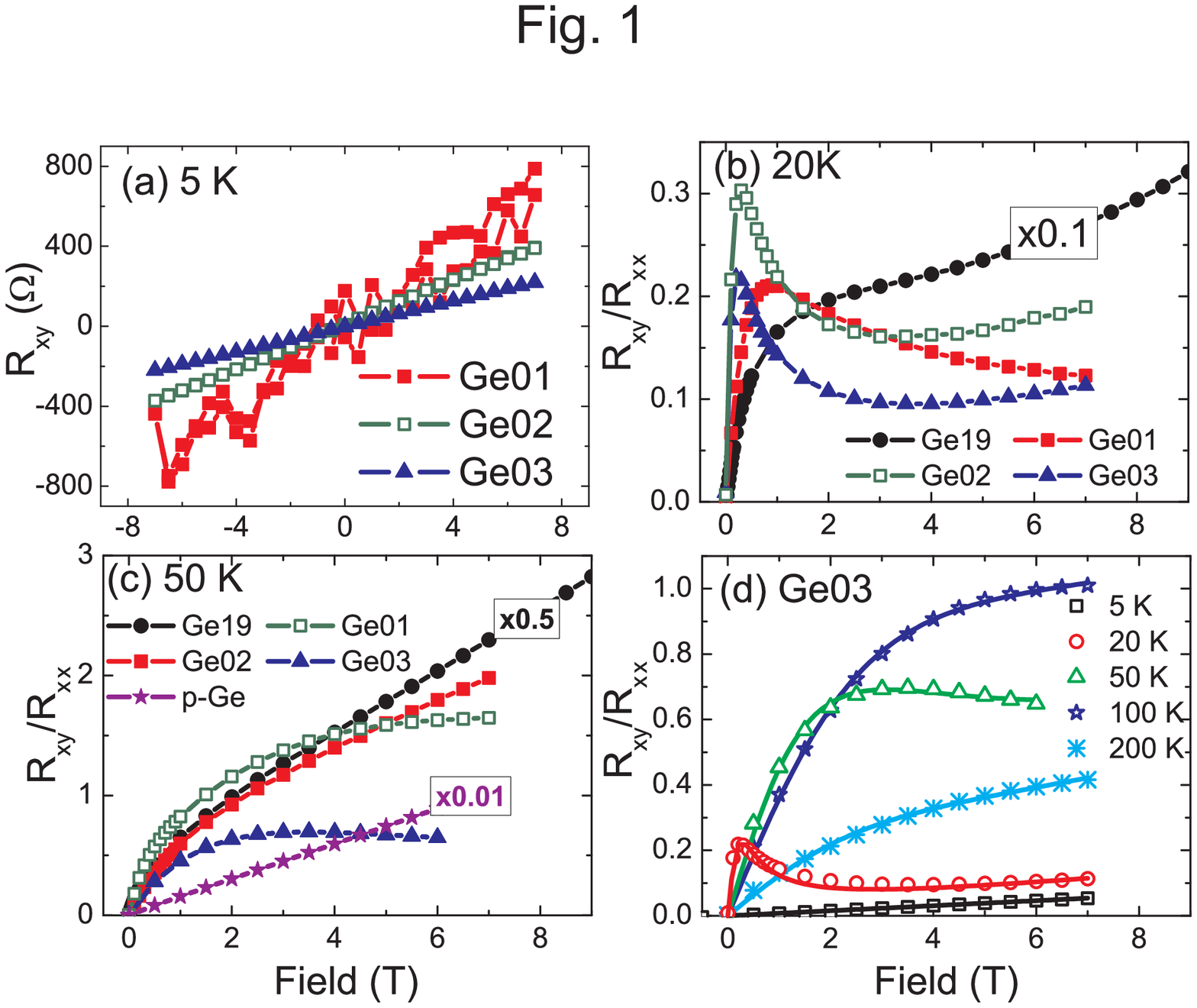}
\end{figure*}

\begin{figure*} \center
\includegraphics[scale=1]{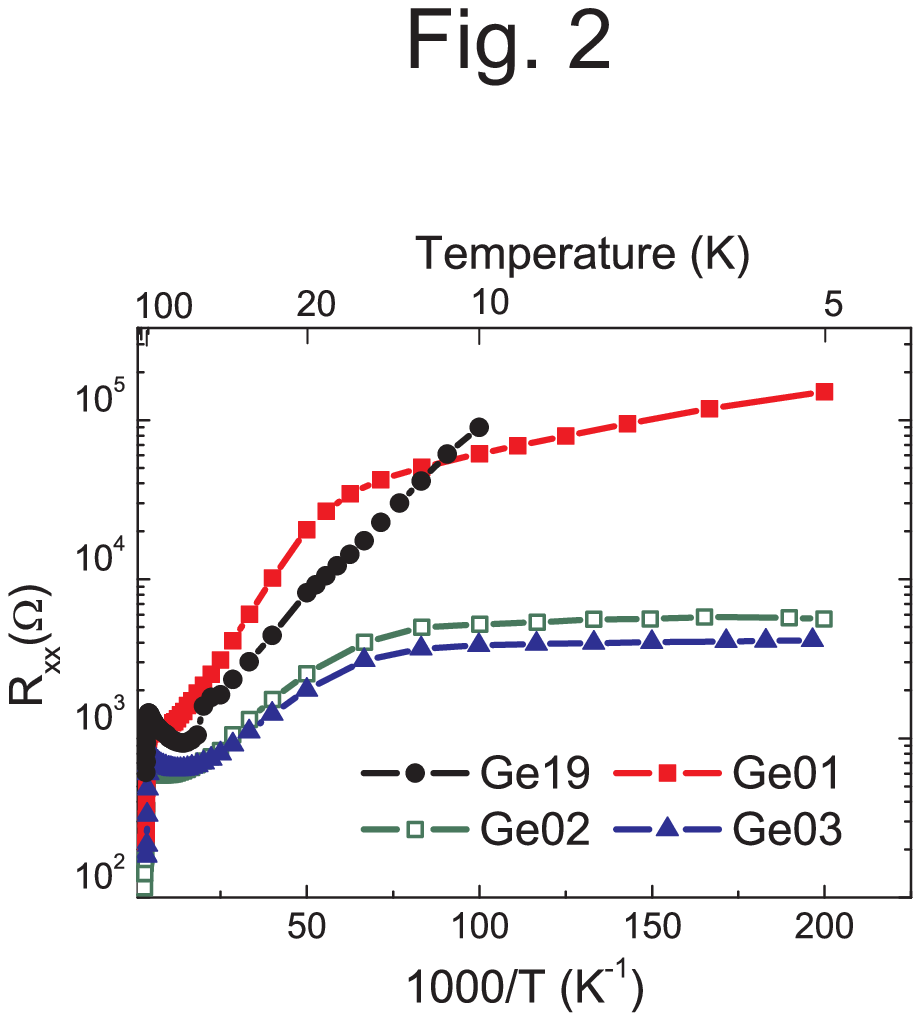}
\end{figure*}

\begin{figure*} \center
\includegraphics[scale=0.8]{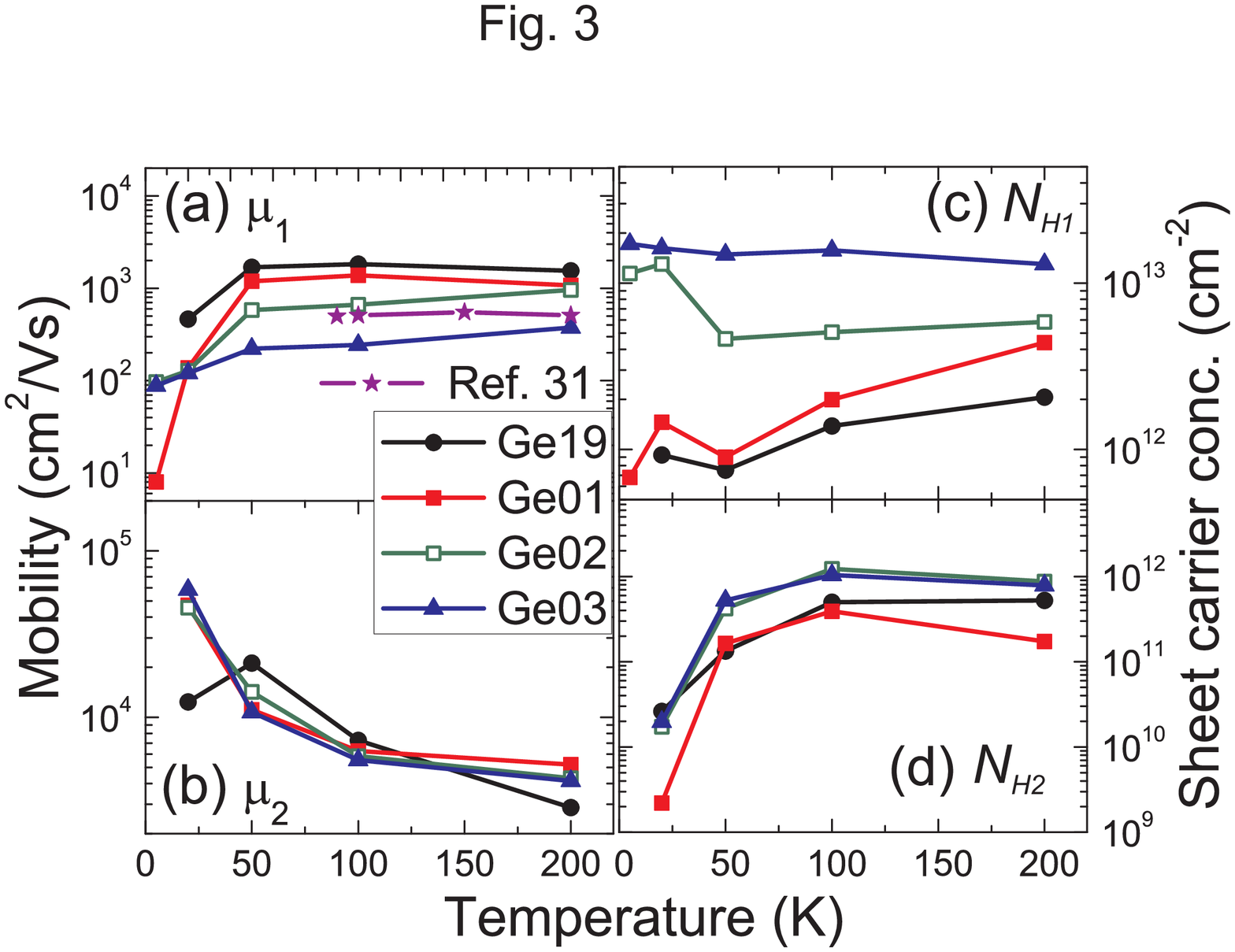}
\end{figure*}

\end{document}